# Universal law for waiting internal time in seismicity and its implication to earthquake network

SUMIYOSHI ABE[1] and NORIKAZU SUZUKI[2]

[1] *Department of Physical Engineering, Mie University, Mie 514-8507, Japan*

[2] *College of Science and Technology, Nihon University, Chiba 274-8501, Japan*

**Abstract** — In their paper (*Europhys. Lett.*, **71** (2005) 1036), Carbone, Sorriso-Valvo, Harabaglia and Guerra showed that "unified scaling law" for conventional waiting times of earthquakes claimed by Bak *et al.* (*Phys. Rev. Lett.*, **88** (2002) 178501) is actually not universal. Here, instead of the conventional time, the concept of the internal time termed the *event time* is considered for seismicity. It is shown that, in contrast to the conventional waiting time, the waiting event time obeys a power law. This implies the existence of temporal long-range correlations in terms of the event time with no sharp decay of the crossover type. The discovered power-law waiting event-time distribution turns out to be universal in the sense that it takes the same form for seismicities in California, Japan and Iran. In particular, the parameters contained in the distribution take the common values in all these geographical regions. An implication of this result to the procedure of constructing earthquake networks is discussed.

______________________________________________________________________

Seismicity is characterized by diverse phenomenological natures including two celebrated classical laws: the Gutenberg-Richter law for the frequency-magnitude scaling and the Omori law for the temporal pattern of aftershocks [1]. Those laws have continuously been attracting physicists' attention, because they show how seismicity is a complex phenomenon [2]. They are however mostly empirical, since microscopic dynamics governing the phenomenon is largely unknown. In such a situation, it is important to clarify the property of correlations. It is known that an earthquake can trigger the next one that can be more than 1000 km away [3], implying that the event-event correlation is long-ranged. In fact, the distribution of the spatial distance between two successive earthquakes obeys a definite law [4]. The interoccurrence time (or, calm time) between two successive events also exhibits a definite statistical behavior [5,6]. Both of them significantly deviate from Poissonian statistics. These specific laws naturally allow one to put a working hypothesis that two successive events are correlated at least at a statistical level. Physically, the long-range event-event correlation is considered to be induced by seismic waves with small values of the wave number.

Recently, we have proposed the network representation of seismicity [7]. There, the event-event correlations are represented by the edges of an earthquake network (see below for details). The topological and statistical properties of an earthquake network give information about the correlations in seismicity in a peculiar manner. This approach turned out to reveal many new aspects of complexity of seismicity. The earthquake networks constructed using the seismic data taken from several different geographical regions such as California and Japan were found to be complex networks [7-10], being scale-free, small-world and hierarchically organized, each of which has its own implication in view of seismology.

In the studies of seismicity, one can consider two different kinds of time: one is the conventional time, and the other is the *internal time*. Let $\{t_1, t_2, ..., t_N\}$ be the conventional occurrence times of $N$ earthquakes contained in the dataset to be analyzed. In this case, the internal time is simply the label "$n$" of $t_n$ ($n = 1, 2, ..., N$), which is henceforth referred to as the "event time". (The "natural time", $\chi_n$, discussed in ref. [11] is the rescaled event time: $\chi_n = n/N$.) In ref. [12], we have discussed the concept of period in the directed earthquake network. There, the period is defined as the path length of a loop, which implies after how many events a given starting vertex is revisited. So, it can be thought of as the *waiting event time*, which is denoted by $n_W$. We have found [12] that the waiting event-time distribution (i.e., the period distribution) obeys a power law. On the other hand, it has been discussed in ref. [13]

that the conventional waiting time distribution for the cells associated with the two-dimensional division of the region of California satisfies a scaling law with respect to the values of the cell size and magnitude. (Here, the conventional waiting time for a certain cell is the ordinary time interval between two succcessive events occurred in that cell.) The conclusion of ref. [13] is that long-time correlations are weak and can be ignored since the scaling function sharply decays for large values of the rescaled conventional waiting time in the crossover type. Here is the important physics concerning the concept of time in the complex seismic phenomenon. That is, the long-time correlations can be observed in the event time and not in the conventional time. In addition, we also note the work in ref. [14] that the scaling law presented in ref. [13] is not universal and does depend on datasets. Therefore, a crucial question to be asked is if the concept of internal time, i.e., event time, can reveal a universal scaling law that cannot be established as long as the conventional time is concerned.

In this paper, we affirmatively answer the above question. We report the discovery of a new law for the waiting event time in seismicity. We show that the law is quite universal in the sense that it takes the same form in three different geographical regions: California, Japan and Iran. Quite remarkably, the data collapses for the waiting event-time distributions in these regions are realized *by the common values of* *parameters contained in the law* (i.e., eqs. (2) and (3) below). No sharp decays of the "scaling functions" occur for large values of the waiting event time, presenting an evidence for the existence of the long-range event-time correlations.

The procedure for constructing an earthquake network presented in ref. [7] is as follows. Firstly, we divide a geographical region under consideration into cubic cells. Secondly, we identify a cell with a vertex of a network if earthquakes occurred therein. Thirdly, we link two vertices by an edge, if they are of two successive events. If two successive events occur in the same cell, then we attach a tadpole (i.e., a self-loop) to that vertex. These edges and tadpoles represent the event-event correlations, in conformity with our working hypothesis mentioned earlier. Using this procedure, we can map a seismic time series to a growing directed stochastic network, which is an earthquake network that we have been referring to. By definition, this network is in a single stroke. Note that the procedure contains a single parameter: the cell size, $L$. We shall examine various values of the cell size.

The concept of period associated with a given vertex in a directed earthquake network is relevant to the question: after how many earthquakes an event returns to the initial vertex. Take a vertex of the network. There are various loops, which start from and end at the taken vertex. Period is then defined as the path length of the loop. For example, let us consider the network in fig. 1 corresponding to a sequence of events:

$\cdots \to v_1 \to v_2 \to v_3 \to v_4 \to v_4 \to v_5 \to v_3 \to v_1 \to v_6 \to v_7 \to v_6 \to v_7 \to v_1 \to v_6 \to \cdots$.

There exist two loops associated with vertex $v_1$: $v_1 \to v_2 \to v_3 \to v_4 \to v_4 \to v_5 \to v_3 \to v_1$ and $v_1 \to v_6 \to v_7 \to v_6 \to v_7 \to v_1$. Their values of period are 7 and 5, respectively. (Note that $v_1 \to v_2 \to v_3 \to v_1$ is not a loop, since it does not belong to the single-stroke network.) These values are those of the waiting event time for vertex $v_1$.

To study the waiting event-time distribution in real seismicity, we have employed three different datasets taken from (i) California (http://www.data.scec.org), (ii) Japan (http://www.hinet.bosai.go.jp) and (iii) Iran (http://irsc.ut.ac.ir). The regions covered are (i) 28.00°N–39.41°N latitude and 112.10°W–123.62°W longitude with the maximal depth 175.99 km, (ii) 17.96°N–49.31°N latitude and 120.12°E–156.05°E longitude with the maximal depth 681.00 km and (iii) 23.89°N–43.51°N latitude and 41.32°E–68.93°E longitude with the maximal depth 36.00 km. The intervals covered are between (i) 00:25:8.58 on January 1, 1984 and 23:15:43.75 on December 31, 2006, (ii) 00:02:29.62 on June 3, 2002 and 23:45:9.57 on December 31, 2007 and (iii) 03:08:11.10 on January 1, 2006 and 18:26:21.90 on December 31, 2008. The total numbers of earthquakes occurred are (i) 404106, (ii) 724535 and (iii) 22845, respectively. Given these intervals, no spatial windeows are set, and all events contained are treated in our analysis. In particular, no artificial manipulations are made on the datasets, in contrast to the work in ref. [13], in which events occuring in the intervals shorter than 38 s are eliminated.

In fig. 2, we present plots of the histograms of the waiting event-time distributions in those three geographical regions for various values of the cell size as well as the threshold of magnitude. There, it is observed that the distributions are of the power-law type with different values of the exponent. We have ascertained that these values of the exponent come from the cell size and *depend on neither the thereshold of magnitude nor the fractal dimensions*, unlike the case of the conventional waiting time distribution considered in ref. [13]. Accordingly, we have examined if the data collapse occurs in terms of various values of the cell size. The result for each region is presented in fig. 3. Here, the cell size, $L$, is made dimensionless in the following way [10]: $l \equiv L/(L_{LAT} L_{LON})^{1/2}$, where $L_{LAT}$ and $L_{LON}$ are the dimensions in the directions of latitude and longitude of the region, respectively. The reason why depth, $L_{DEP}$, is not considered is due to the fact that most of the events occurred in the shallow regions, i.e., the dimension of depth is much smaller than those of latitude and longitude. [In fact, the values of $(L_{LAT}, L_{LON}, L_{DEP})$ covered by the datasets are: (i) (1268.98 km, 1055.83 km, 175.99 km), (ii) (3485.30 km, 3239.70 km, 681.00 km) and (iii) (2181.30 km, 2529.29

km, 36.00 km), but in all three regions, more than 90% of the events are shallower than

(i) 13.86 km, (ii) 69.00 km and (iii) 26.60 km.] As can be seen in fig. 3, the waiting

event-time distribution, $P(n_W, l)$, satisfies the following relation:

$$P(n_W, l) \sim (n_W)^{-\gamma(l)}, \tag{1}$$

where

$$\gamma(l) = c\, l^\alpha. \tag{2}$$

This result on the waiting event-time distribution should be compared with that on the conventional waiting time distribution in ref. [13]. First of all, in marked contrast with the assertion in ref. [13], there is no sharp decay of the crossover type for large values of the waiting event time, demonstrating the existence of long event-time correlations.

Secondly, the discovered law in eq. (1) has high universality in the sense that $c$ and $\alpha$

in eq. (2) are, quite remarkably,

$$c \cong 2.50, \qquad \alpha \cong 0.14, \tag{3}$$

*for all three independent datasets.* We note that the data collapse itself is established irrelevantly to the value of *c*. To make the universality of our discovery more explicit, we present in figs. 4 and 5 the waiting event-time distributions in all those regions. In fig. 5, one sees how the data collapse of the bare plots in fig. 4 can be established based on the model in eq. (1) with eqs. (2) and (3). To quantify the quality of the data collapse, we have evaluated the ratio, $W^{(5)}_{max} / W^{(4)}_{max}$, where $W^{(4)}_{max}$ and $W^{(5)}_{max}$ are the maximum values of the vertical width of the distributed points in figs. 4 and 5, respectively. The calculated value is $W^{(5)}_{max} / W^{(4)}_{max} = 1.1 \times 10^{1} / 4.7 \times 10^{4} \cong 2.3 \times 10^{-4}$, showing the good quality of the data collapse.

In conclusion, we have studied the internal time termed the event time in seismicity instead of the conventional time, and have discovered a new law for the waiting event time. This law turned out to be universal in the sense that it takes the same form for seismicities in California, Japan and Iran with the common values of the parameters, *c* and $\alpha$, as in eq. (3). In particular, the result confirms the existence of long event-time correlations in seismicity, which cannot be observed as long as the conventional time is employed.

The present result has an important implication to the concept of earthquake network. The procedure of constructing an earthquake network orginally proposed in ref. [7] is directly concerned with the event time, the lapse of which corrresponds to the path lenghth between two relevant edges of a directed earthquake network. The networks thus constructed from real data are known to possess a series of highly universal properties [7-10]. On the other hand, there is yet another procedure of constructing an earthquake network [15] that is guided by the work in ref. [13]. It contains a number of parameters, but universal features are hardly identified. The present work puts a basis on the reason why the procedure proposed in ref. [7] is plausible.

$$* \quad * \quad *$$

SA was supported in part by a Grant-in-Aid for Scientific Research from the Japan Society for the Promotion of Science.

# Figure Caption

Fig. 1    A schematic description of a directed earthquake network.

Fig. 2    Plots of the waiting event-time distribution in (i) California, (ii) Japan and (iii) Iran, for various values of the threshold of magnitude, $M_{th}$, and the cell size, $L$ [km]. Both $n_W$ and $P(n_W)$ are dimensionless.

Fig. 3    Data collapse of fig. 2 based on the model in eq. (1) with eqs. (2) and (3).

Fig. 4    (color online) Plots of the waiting event-time distributions in three regions in fig. 2 (C: California, J: Japan and I: Iran) are gathered.

Fig. 5    (color online) Data collapse of fig. 4 based on the model in eq. (1) with eqs. (2) and (3).

Fig. 1

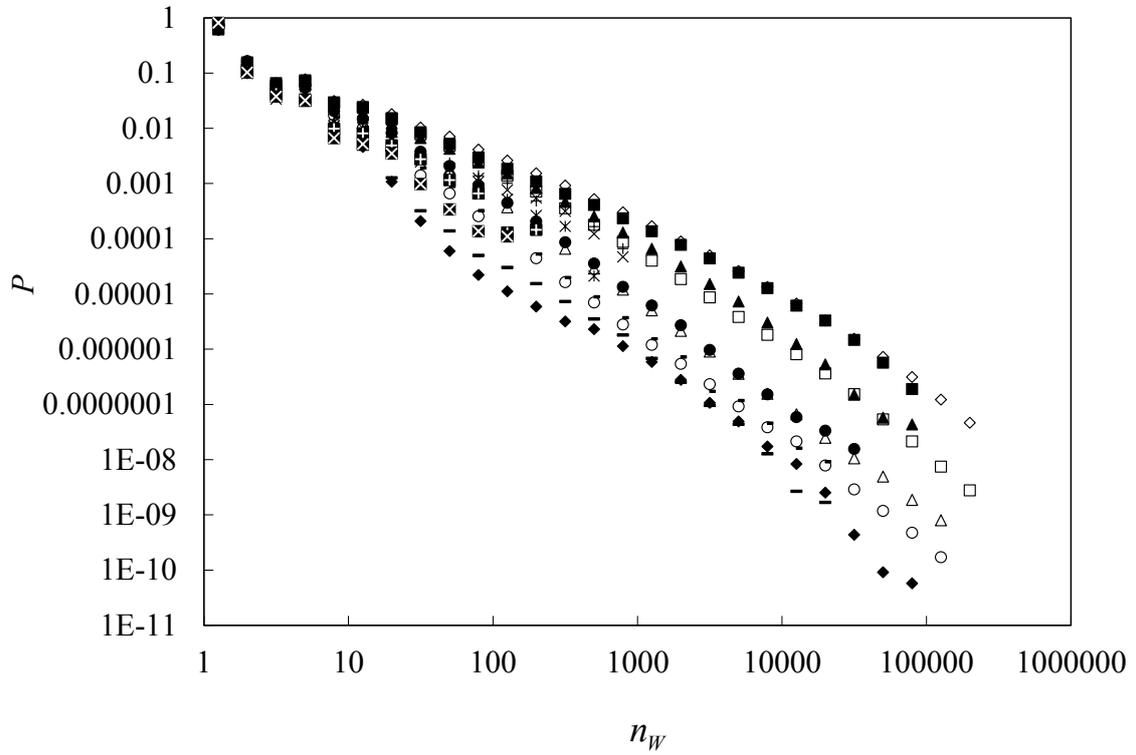

$(M_{th}, L)$

| | | | | |
|---|---|---|---|---|
| ◇ (no, 10) | □ (no, 30) | △ (no, 100) | ○ (no, 200) | ◆ (no, 400) |
| ■ (2, 10) | ▲ (2, 30) | ● (2, 100) | ‐ (2, 200) | — (2, 400) |
| + (4, 10) | × (4, 30) | ✳ (4, 100) | ⊞ (4, 200) | ⊠ (4, 400) |

Fig. 2(i)

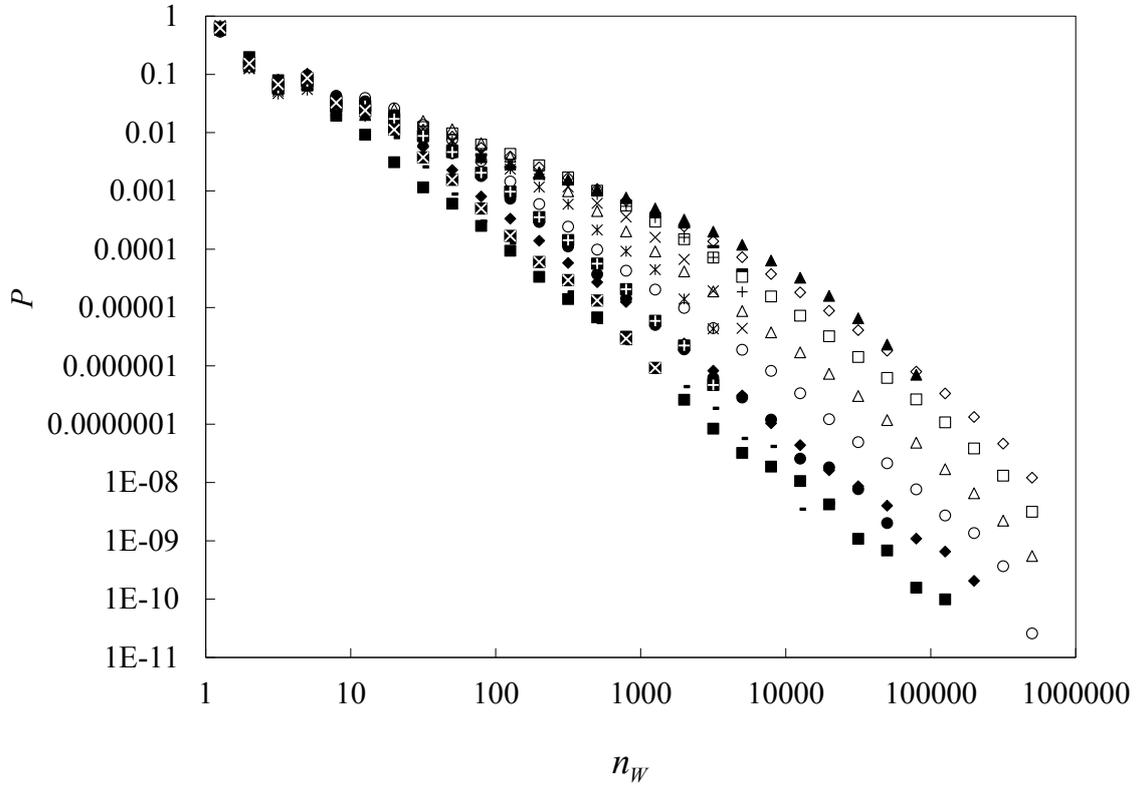

$(M_{\text{th}}, L)$

| | | | | |
|---|---|---|---|---|
| ◇ (no, 30) | □ (no, 50) | △ (no, 100) | ○ (no, 200) | ◆ (no, 400) |
| ■ (no, 800) | ▲ (2, 30) | ● (2, 400) | ▬ (2, 800) | ▬ (4, 30) |
| + (4, 50) | × (4, 100) | ✹ (4, 200) | ✚ (4, 400) | ⊠ (4, 800) |

Fig. 2(ii)

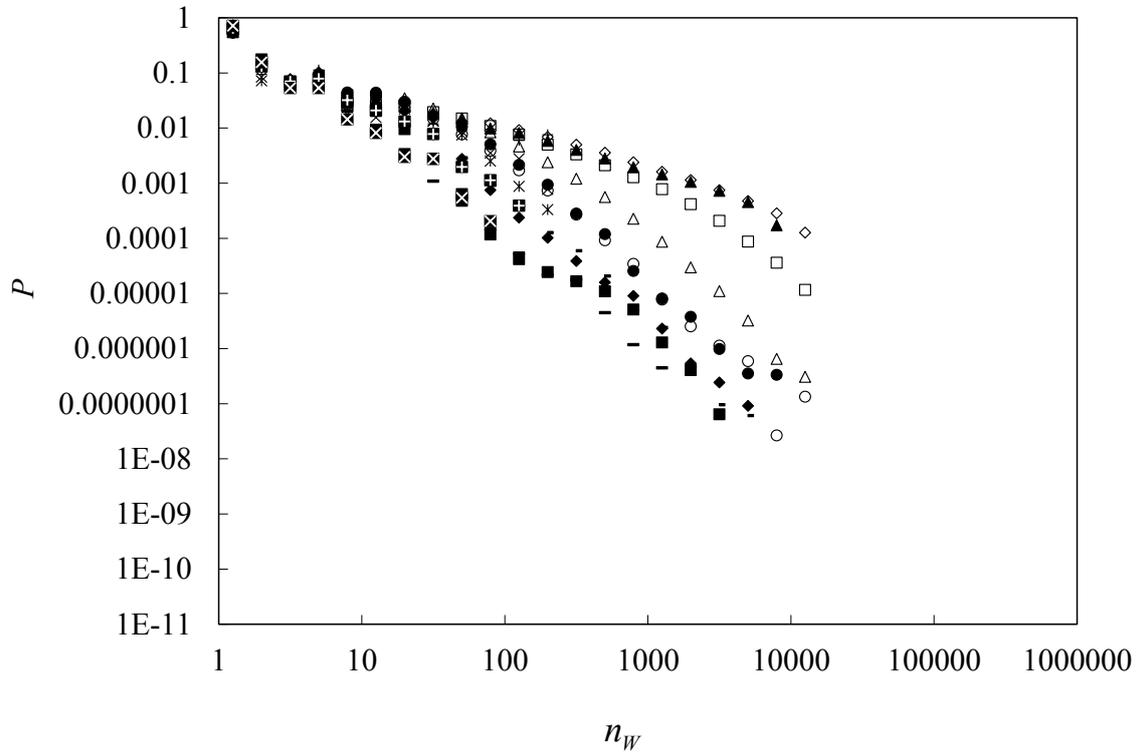

$(M_{th}, L)$

| | | | | |
|---|---|---|---|---|
| ◇ (no, 10) | □ (no, 30) | △ (no, 100) | ○ (no, 200) | ◆ (no, 400) |
| ■ (no, 800) | ▲ (2, 10) | ● (2, 200) | ‐ (2, 400) | ― (2, 800) |
| + (4, 10) | × (4, 100) | ✳ (4, 200) | ✚ (4, 400) | ⊠ (4, 800) |

Fig. 2(iii)

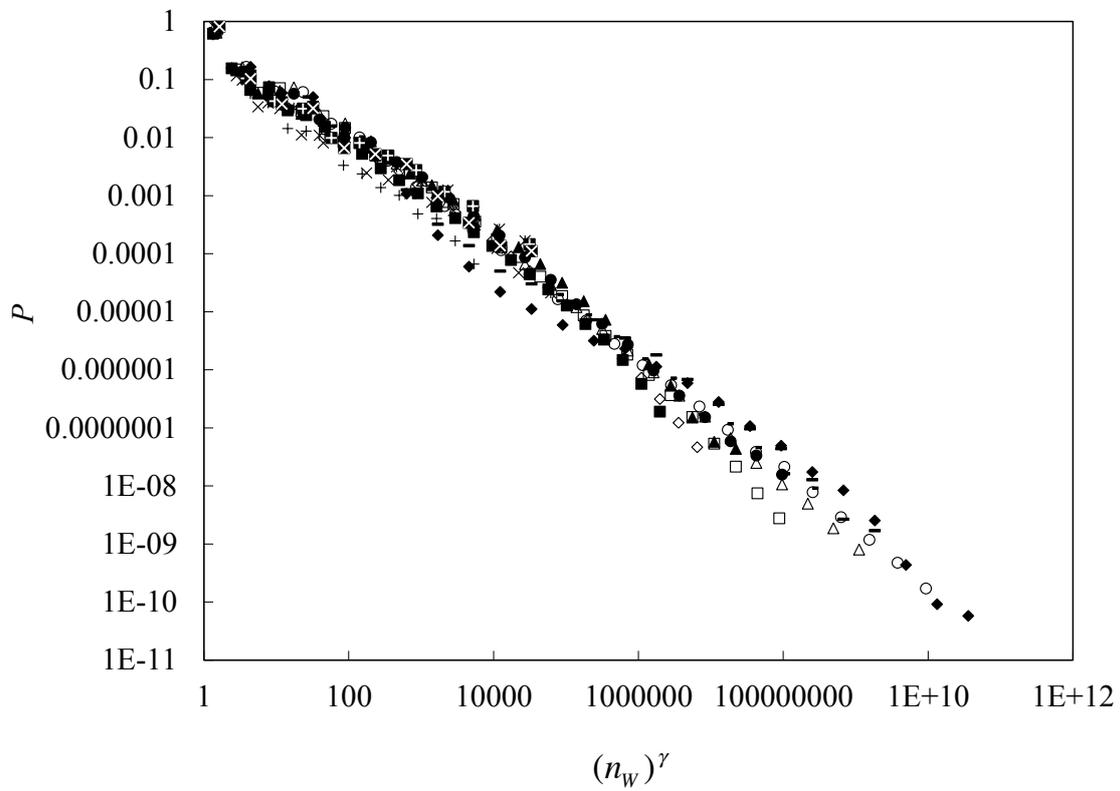

$(M_{th}, L)$

| | | | | |
|---|---|---|---|---|
| ◇ (no, 10) | □ (no, 30) | △ (no, 100) | ○ (no, 200) | ◆ (no, 400) |
| ■ (2, 10) | ▲ (2, 30) | ● (2, 100) | - (2, 200) | — (2, 400) |
| + (4, 10) | × (4, 30) | ✻ (4, 100) | ✜ (4, 200) | ⊠ (4, 400) |

Fig. 3(i)

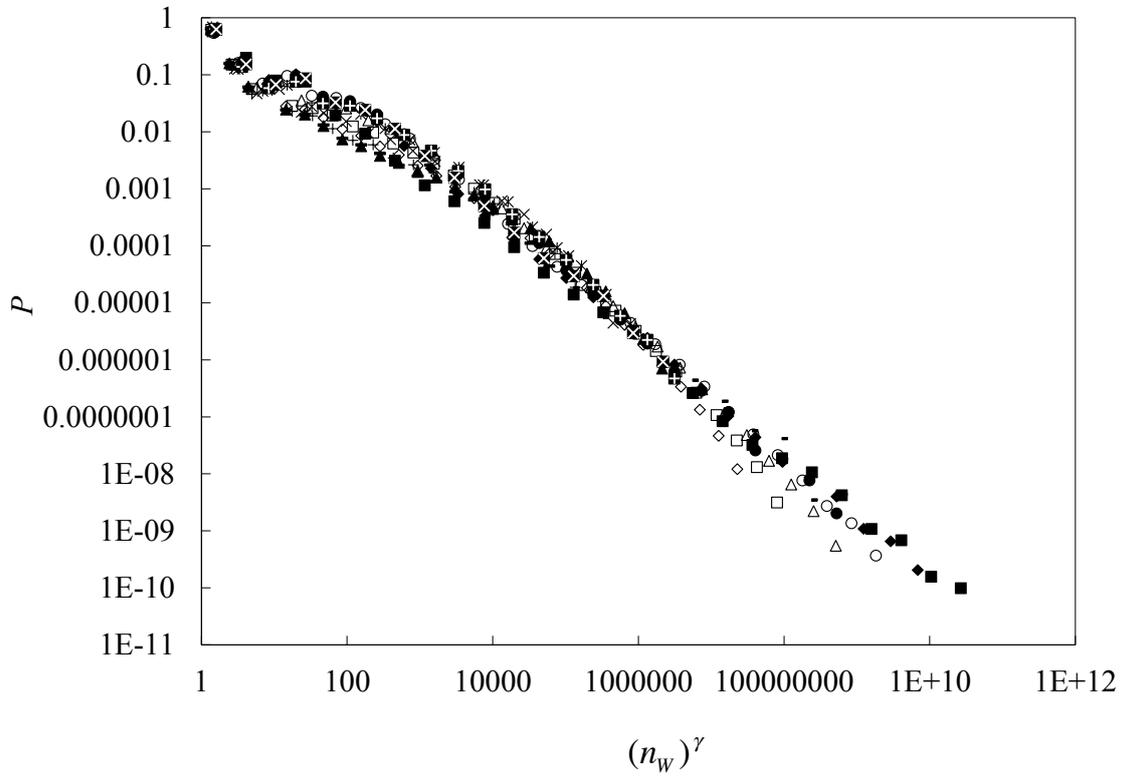

Fig. 3(ii)

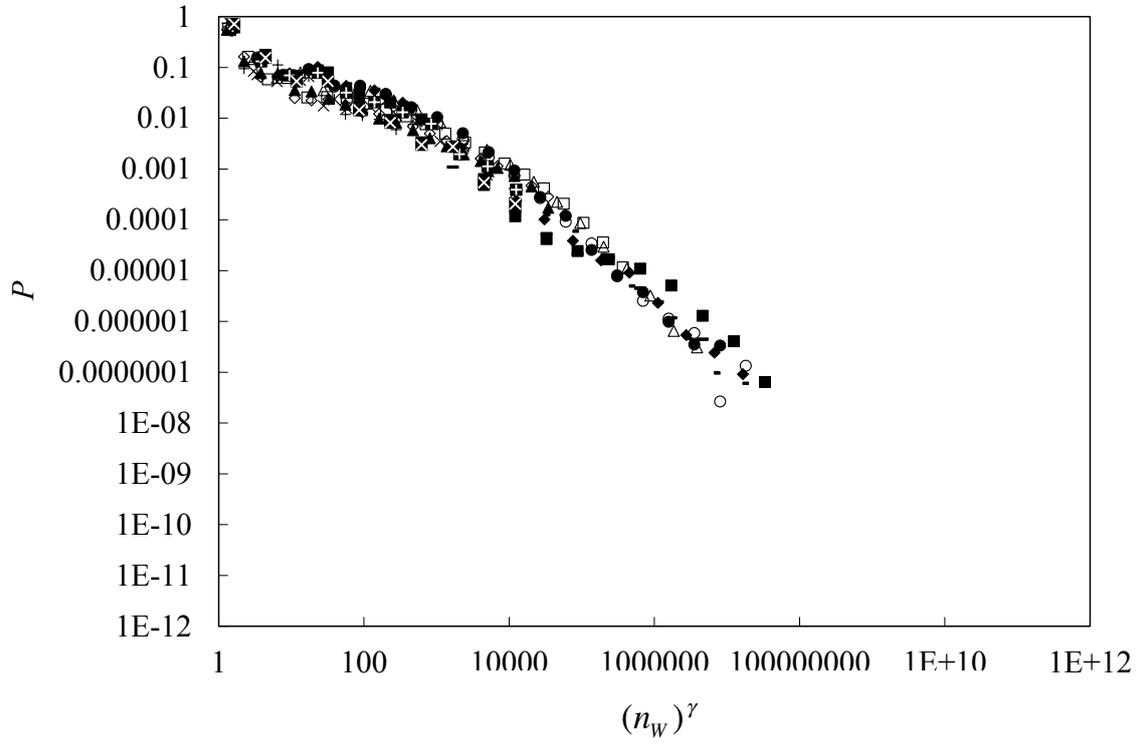

$(M_{th}, L)$

| | | | | |
|---|---|---|---|---|
| ◇ (no, 10) | □ (no, 30) | △ (no, 100) | ○ (no, 200) | ◆ (no, 400) |
| ■ (no, 800) | ▲ (2, 10) | ● (2, 200) | - (2, 400) | ▬ (2, 800) |
| + (4, 10) | × (4, 100) | ✱ (4, 200) | ✜ (4, 400) | ⊠ (4, 800) |

Fig. 3(iii)

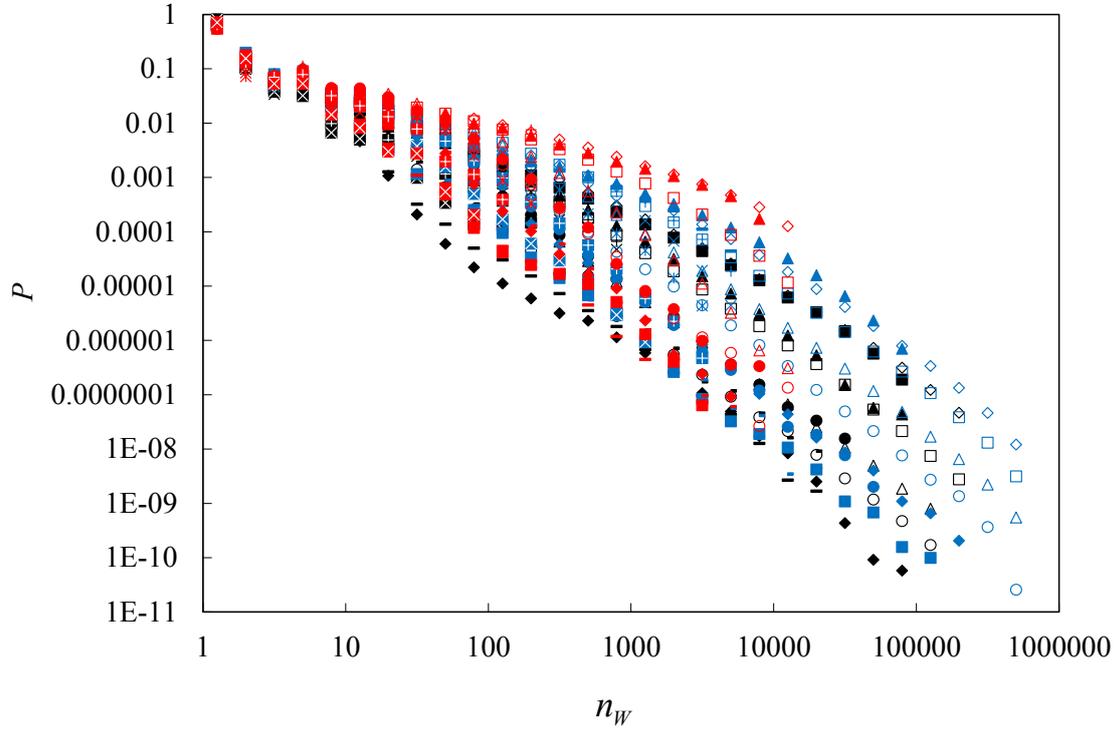

(region, $M_{th}$, $L$)

| | | | | |
|---|---|---|---|---|
| ◇(C, no, 10) | □(C, no, 30) | △(C, no, 100) | ○(C, no, 200) | ◆(C, no, 400) |
| ■(C, 2, 10) | ▲(C, 2, 30) | ●(C, 2, 100) | −(C, 2, 200) | −(C, 2, 400) |
| ✚(C, 4, 10) | ✕(C, 4, 30) | ✶(C, 4, 100) | ⊞(C, 4, 200) | ⊠(C, 4, 400) |
| ◇(J, no, 30) | □(J, no, 50) | △(J, no, 100) | ○(J, no, 200) | ◆(J, no, 400) |
| ■(J, no, 800) | ▲(J, 2, 30) | ●(J, 2, 400) | −(J, 2, 800) | −(J, 4, 30) |
| ✚(J, 4, 50) | ✕(J, 4, 100) | ✶(J, 4, 200) | ⊞(J, 4, 400) | ⊠(J, 4, 800) |
| ◇(I, no, 10) | □(I, no, 30) | △(I, no, 100) | ○(I, no, 200) | ◆(I, no, 400) |
| ■(I, no, 800) | ▲(I, 2, 10) | ●(I, 2, 200) | −(I, 2, 400) | −(I, 2, 800) |
| ✚(I, 4, 10) | ✕(I, 4, 100) | ✶(I, 4, 200) | ⊞(I, 4, 400) | ⊠(I, 4, 800) |

Fig. 4

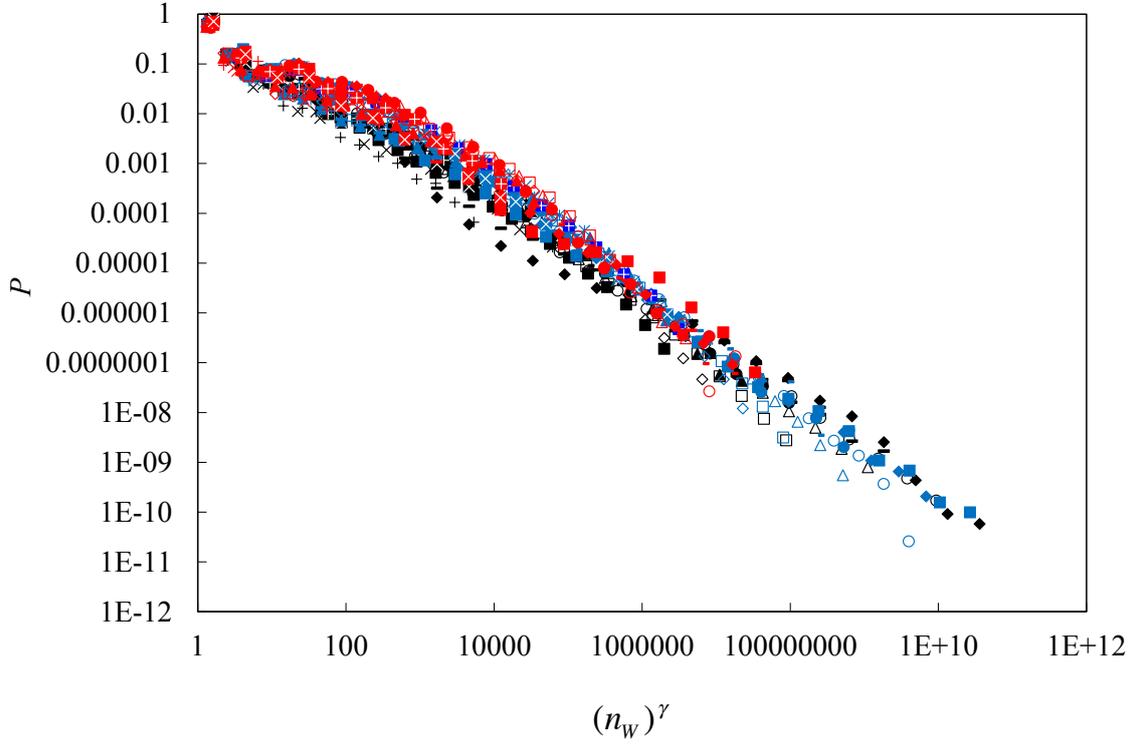

| (region, $M_{th}$, $L$) | | | | |
|---|---|---|---|---|
| ◇(C, no, 10) | □(C, no, 30) | △(C, no, 100) | ○(C, no, 200) | ◆(C, no, 400) |
| ■(C, 2, 10) | ▲(C, 2, 30) | ●(C, 2, 100) | -(C, 2, 200) | —(C, 2, 400) |
| +(C, 4, 10) | ✕(C, 4, 30) | ✳(C, 4, 100) | ⊞(C, 4, 200) | ⊠(C, 4, 400) |
| ◇(J, no, 30) | □(J, no, 50) | △(J, no, 100) | ○(J, no, 200) | ◆(J, no, 400) |
| ■(J, no, 800) | ▲(J, 2, 30) | ●(J, 2, 400) | -(J, 2, 800) | —(J, 4, 30) |
| +(J, 4, 50) | ✕(J, 4, 100) | ✳(J, 4, 200) | ⊞(J, 4, 400) | ⊠(J, 4, 800) |
| ◇(I, no, 10) | □(I, no, 30) | △(I, no, 100) | ○(I, no, 200) | ◆(I, no, 400) |
| ■(I, no, 800) | ▲(I, 2, 10) | ●(I, 2, 200) | -(I, 2, 400) | —(I, 2, 800) |
| +(I, 4, 10) | ✕(I, 4, 100) | ✳(I, 4, 200) | ⊞(I, 4, 400) | ⊠(I, 4, 800) |

Fig. 5